\documentclass[a4paper,12pt,oneside,reqno]{article}
\usepackage[utf8x]{inputenc}
\usepackage{authblk}
\usepackage{amsmath}
\usepackage[pdfborder={0 0 0}]{hyperref}
\usepackage{cite}

\title{Nucleation stage in supersaturated vapor with inhomogeneities due to nonstationary diffusion onto growing droplets.}
\date{}

\author[1]{Anatoly Kuchma}
\author[1,2]{Maxim Markov}
\author[1]{Alexander Shchekin\thanks{Corresponding author, \textit{e-mail address}: \texttt{akshch@list.ru} }}

\affil[1]{Department of Statistical Physics, Faculty of Physics\\ St Petersburg State University, Ulyanovskaya 1, Petrodvoretz\\  St Petersburg, 198504, Russian Federation}
\affil[2]{Laboratoire des Solides Irradies, Ecole Polytechnique, 91128 Palaiseau Cedex, France}

\newcommand{\Keywords}[1]{\par\noindent
{\small{\em Keywords\/}: #1}}

\newcommand{\ignore}[1]{}

\begin{document}

\maketitle

\begin{abstract}
An analytical description of nucleation stage in a supersaturated vapor with instantly created supersaturation is given with taking into account the vapor
concentration inhomogeneities arising as a result of depletion due to non-stationary diffusion onto growing droplets. This description suggests that the intensity
of the nucleation of new droplets is suppressed in spherical diffusion regions of a certain size surrounding previously nucleated droplets, and remains at the initial
level in the remaining volume of the vapor-gas medium. The value of volume excluded from nucleation depends on the explicit form of the vapor concentration profile in the space
around the growing droplet, and we use for that the unsteady self-similar solution of time-dependent diffusion equation with a convective term describing the flow of
the gas-vapor mixture caused by moving surface of single growing droplet. The main characteristics of the phase transition at the end of the nucleation stage are found
and compared with those in the theory of nucleation with homogeneous vapor consumption (the theory of mean-field vapor supersaturation).
\\
\end{abstract}

\section[]{Introduction.}
The traditional approach used to describe kinetics of the liquid phase formation at the nucleation stage (the stage of nucleation of overcritical, i.e., steadily
growing, droplets), assumes that the consumption of vapor by growing droplets leads to a simultaneous and uniform-over-volume (for the whole ensemble of droplets)
decrease of the vapor supersaturation \cite{ref-1,ref-2,ref-3,ref-4,ref-5}. It should be noted that the approximation of a homogeneous vapor supersaturation in the system (the mean-field
supersaturation) can be justified only on the assumption that the sizes of diffusion vapor shells (from which the droplets consume the vapor) surrounding growing
droplets are large not only in comparison with the sizes of the droplets themselves, but also with respect to the average distance between the droplets. This condition
of diffusion mixing, however, is certainly not satisfied at the stage of nucleation of overcritical droplets, when the average distance between the droplets can be
very large. At non-stationary diffusion to a droplet, the corresponding diffusion shell in the vapor-gas medium is localized around the droplet, and such shells for
neighbour droplets may overlap only to the end of the nucleation stage. At the same time, the nucleation intensity is exponentially sensitive to the local vapor
supersaturation. As a consequence, the problem to develop a description of the nucleation kinetics, which would explicitly take into account the heterogeneity of the
field of vapor concentration caused by vapor diffusion onto consuming droplets, becomes an actual one.

The effect of vapor heterogeneity caused by vapor diffusion to a growing drop had previously been considered in several works with using different approaches 
\cite{ref-6,ref-7,ref-8,ref-9,ref-10,ref-11}. In particular, this problem was raised in Refs.\cite{ref-6,ref-7} where the average nucleation rate per unit volume of
vapor-gas medium notion of “clearance volume” around the droplet was introduced. It was noted by Kurasov in Refs.\cite{ref-8,ref-9} that using the stationary diffusion
approximation for the density profile around the droplet is not appropriate, and the non-stationary density profile should be taken into account with recognizing 
the presence of a shell around the droplet where nucleation of new droplets is absent. It was proposed an integral equation for the available for nucleation vapor
volume with calculation of the vapor profile around droplets in the approximation of non-moving droplet surface. Grinin et al \cite{ref-10,ref-11} calculated the probability of 
the nearest-neighbor droplet nucleation in the diffusion profile of a previously nucleated droplet found under condition that material balance maintains between the
droplet and the vapor, even though the droplet boundary is a moving one. Their approach operated with the mean distance to the nearest-neighbor droplet and the mean time
to its appearance which provide estimates for the duration of the nucleation stage and the number of droplets formed per unit volume during the nucleation stage.

In this paper, we assume as well as in Refs.\cite{ref-6,ref-7,ref-8,ref-9,ref-10,ref-11} that the intensity of nucleation of new droplets is suppressed in spherical 
diffusion layers of certain thickness surrounding the previously nucleated drops, and stays at the initial level in the remaining volume of the gas-vapor mixture. It
can be called the excluded volume approach because the total volume of such diffusion spheres with droplets at their centers is excluded from the nucleation process. The
value of the excluded volume depends on the explicit form of the vapor concentration profile in the space around the growing droplet, and we use for that the most
accurate unsteady self-similar solution of time-dependent diffusion equation with a convective term describing the flow of the gas-vapor mixture caused by moving
surface of the growing droplet. Earlier, the similar approach was used to describe the stage of nucleation of bubbles in gas-supersaturated solution \cite{ref-12,ref-13}.
However it should be noted that the physical appearance of the “excluding” mechanism is different for growing bubbles and droplets, and the differences are outlined below.
As a part of excluded volume approach for droplets, we calculate the main characteristics of the phase transition (the duration of the stage, the number of nucleated
droplets per unit volume, the maximal and average radii of the droplets) set at the end of the nucleation stage, and the results are compared with the results of the
theory of mean field of vapor supersaturation.

The article is organized as following. First, in Section 2 we reformulate the basics of the theory of the nucleation stage in the approximation of the
mean-field vapor concentration in the convenient for subsequent comparison form. Description of nucleation in the framework of the excluded volume approach is
constructed in Section 3. In Section 4 we compare the results from Sections 2 and 3 and make conclusions in Section 5.

\section[The mean-field approximation.]{Nucleation stage in the mean-field approximation for vapor concentration.}
Let us formulate the basic principles of the theory of nucleation stage at instantly created vapor supersaturation in the mean-field approximation for vapor
concentration \cite{ref-1,ref-2,ref-3,ref-4,ref-5}. The main characteristics of the state of droplets, reached to the end of the nucleation stage are: the total number
of overcritical droplets, their maximum and average sizes, the time of development of the nucleation stage. The theoretical problem is to link these main
characteristics of the final state of droplets to initial value of supersaturation in the system. Here we assume a uniform vapor consumption by growing droplets.
We restrict ourselves to the case of isothermal nucleation at which the temperature of droplets equals to the vapor-gas medium temperature. 

The number \textit{I} (the nucleation rate) of overcritical droplets emerging at a given vapor supersaturation per unit time per unit volume of the gas-vapor system is
given as \cite{ref-1}
\begin{equation} \label{eq:1}
I(\zeta) = A(\zeta)\exp(-\Delta F(\zeta))
\end{equation}
where $\zeta(t) = (n(t) - n_{\infty})/n_{\infty}$ is the vapor supersaturation at the time $t$, $n(t)$ is the number density of the vapor molecules at time $t$, 
$n_{\infty}$ is the equilibrium number density of vapor molecules saturated with a plane liquid-vapor interface, $\Delta F(\zeta)$ is the work of critical droplet
formation expressed in terms of thermal units $k_{B}T$, $k_{B}$ is the Boltzmann constant, $T$ is the absolute temperature of the system, factor $A(\zeta)$
(the Zel'dovich factor) is much more slowly varying function of vapor supersaturation to compare with exponential factor $\exp(-\Delta F(\zeta))$.

Because of the extremely sharp exponential dependence of the nucleation rate $I$ on vapor supersaturation, reducing the supersaturation by value of a few
percent of the initial supersaturation $\zeta_{0}$ leads to a significant drop in the nucleation rate. As a consequence, we can assume that the relative decline 
$\varphi$ of vapor supersaturation, defined by the expression
\begin{equation} \label{eq:2}
\varphi(t) = \frac{\zeta_{0} - \zeta(t)}{\zeta_{0}}
\end{equation}
satisfies strong inequality
\begin{equation} \label{eq:3}
\varphi \ll 1
\end{equation}
throughout the whole nucleation stage. In view of Eq.\eqref{eq:3}, expression \eqref{eq:1} for the nucleation rate can be reduced with a high degree of accuracy to
expression
\begin{equation} \label{eq:4} 
I(\zeta) = I_{0}\exp(-\Gamma\varphi)
\end{equation}
where $I_{0}=I(\zeta)_{0}$ and the parameter $\Gamma$ are determined by
\begin{equation} \label{eq:5}
\Gamma \equiv -\zeta_{0}\left(\frac{d \Delta F(\zeta)}{d\zeta}\right)_{\zeta=\zeta_{0}}
\end{equation}
The parameter $\Gamma$ appears to be approximately equal to the number of molecules in the droplet of the critical size, i.e., $\Gamma>40$. The fact that 
$\Gamma \gg 1$ allows us to keep in Eq. \eqref{eq:4} only the first two terms of the Taylor expansion of $\Delta F(\zeta)$ with respect to the initial supersaturation
$\zeta_{0}$.

An explicit expression for the work of formation of a critical homogeneous droplet at temperature $T$ is \cite{ref-3}
\begin{equation} \label{eq:6}
\Delta F(\zeta) = \frac{4}{27} \left(\frac{4\pi\sigma}{k_{B}T}\right)^{3}\left(\frac{3}{4\pi n_{l}}\right)^{2}\frac{1}{\ln^{2}(1+\zeta)}
\end{equation}
where $\sigma$ is the droplet surface tension, $n_{l}$ is the number density of molecules in the liquid phase. Substituting Eq.\eqref{eq:6} in Eq.\eqref{eq:5} determines parameter 
$\Gamma$ as a function of initial supersaturation $\zeta_{0}$,
\begin{equation} \label{eq:7}
\Gamma = \frac{8}{27} \left(\frac{4\pi\sigma}{k_{B}T}\right)^{3}\left(\frac{3}{4\pi n_{l}}\right)^{2}\frac{\zeta_{0}}{(1+\zeta_{0})}\frac{1}{\ln^{3}(1+\zeta_{0})}
\end{equation}
For example, for water at $T = 270$ K we have $n_{l} = 3.3\cdot10^{22}$ cm$^{-3}$, $n_{\infty} = 1.3\cdot10^{17}$ cm$^{-3}$, $\sigma = 75.44$ dyn/cm  and it follows
from Eq. \eqref{eq:7} that $\Gamma = 48.75$ at $\zeta_{0} = 4$. As can be seen from Eq. \eqref{eq:4}, the assumption of smallness of the relative decline $\varphi$
during the nucleation stage is justified in this case. 

Let us restrict discussion to a situation where the flow of vapor onto the growing overcritical droplet is controlled by diffusion (during the whole nucleation stage),
so that the inequality $R \gg \lambda$ fulfills where $R$ is the radius of the droplet and $\lambda$ is the mean free path of the molecules in the vapor-gas medium. 
The rate of steady diffusion growth of squared radius $R^{2}$ of markedly overcritical droplet obeys the Maxwell equation \cite{ref-15} 
\begin{equation} \label{eq:8}
\frac{dR^{2}}{dt} = 2D\frac{n_{\infty}}{n_{l}}\zeta_{t}
\end{equation}
where $D$ is the diffusivity of vapor molecules in the vapor-gas medium. Because the rate $dR^{2}/dt$ does not depend on droplet radius and is the same for droplets
growing at supersaturation $\zeta(t)$ at time $t$, then distribution of overcritical droplets in squared radii is given by
\begin{equation} \label{eq:9}
f(R^{2},t) = \int_{0}^{t} d\tau I(\zeta(\tau))\delta(R^{2}-R^{2}(t,\tau))
\end{equation}
where $\delta(x)$ is the Dirac delta function, $R(t,\tau)$ determines at time $t$ the current radius of the droplet nucleated at time $\tau$.

Assuming the radius of the overcritical droplet at the moment of its appearance in the system to be equal to zero, and using Eq. \eqref{eq:8}, we can write 
\begin{equation} \label{eq:10}
R^{2}(t,\tau) = \int^{t}_{\tau}\frac{dR^{2}}{d\widetilde{t}}d\widetilde{t} = 2D\frac{n_{\infty}}{n_{l}}\int_{\tau}^{t}\zeta(\widetilde{t})d\widetilde{t}
\end{equation}
Using Eq. \eqref{eq:2} and definition $\zeta_{0} = (n_{0} - n_{\infty})/n_{\infty}$ ($n_{0}$ is the initial density of the vapor molecules) in Eq. \eqref{eq:10} 
yields
\begin{equation} \label{eq:11}
R^{2}(t,\tau) = 2Da\int_{\tau}^{t}(1-\varphi(\widetilde{t}))d\widetilde{t}
\end{equation}
with parameter $a$ defined as
\begin{equation} \label{eq:12}
a = \frac{n_{\infty}}{n_{l}}\zeta_{0} = \frac {n_{0}-n_{\infty}}{n_{l}} 
\end{equation}
Since, according to Eq. \eqref{eq:3}, current relative decline $\varphi(t)$ of vapor supersaturation is small throughout the nucleation stage, one can neglect it in the
integrand in Eq. \eqref{eq:11}. Finally we have from Eq. \eqref{eq:11}
\begin{equation} \label{eq:13}
R^{2}(t,\tau) = 2Da(t-\tau)
\end{equation}
Thus lowering the mean value of supersaturation on nucleation stage results in the mean-field approximation approach only in a very rapid decrease in the intensity 
of the nucleation and does not affect the rate of growth of the nucleating droplets.

Substituting the expression for $R^{2}(t,\tau)$ in Eq. \eqref{eq:9} and integrating over $\tau$, we obtain an expression for the distribution function $f(R^{2},t)$ through
the function $\varphi (t)$ in the form
\begin{equation} \label{eq:14}
f(R^{2},t) = \frac{I_{0}}{2Da}\exp\left[-\Gamma\varphi\left(t-\frac{R^{2}}{2Da}\right)\right] 
\end{equation}
In its turn, the relative decline $\varphi(t)$ of vapor supersaturation is provided by diffusion transfer to droplets and can be tied with the distribution function with the
help of material balance equations. In a closed system, the corresponding material balance equation can be written as
\begin{equation} \label{eq:15}
(n_{0}-n_{\infty})\varphi(t) = \frac{4\pi}{3}n_{l}\int_{0}^{R^{2}(t)}R^{3}f(R^{2},t)dR^{2}
\end{equation}
where squared radius $R^{2}(t)\equiv R^{2}(t,0) = 2Dat$ corresponds to the maximum size of the droplets in the distribution. Substituting Eq. \eqref{eq:14} into 
Eq. \eqref{eq:15} and passing to the new variable of integration $z \equiv t - R^{2}/2Da$, we obtain the equation for the function $\varphi(t)$ in the form
\begin{equation} \label{eq:16}
\varphi(t) = \frac{8\pi I_{0} D^{3/2}(2a)^{1/2}}{3}\int_{0}^{t}\exp [-\Gamma\varphi(z)](t-z)^{3/2}dz
\end{equation}
The solution of Eq.\eqref{eq:16} can be found with the help of an iterative process, with the zeroth approximation $\varphi=0$. Bearing in mind the small $\varphi(t)$
value on the nucleation stage, we can restrict ourselves by the first iteration. Substituting $\varphi = 0$ in the integrand in \eqref{eq:16}, we obtain
\begin{equation} \label{eq:17}
\varphi(t) = 2^{5/2}\frac{4\pi a^{1/2}}{15}I_{0}D^{3/2}t^{5/2}
\end{equation}
Substituting Eq.\eqref{eq:17} into Eq.\eqref{eq:14} leads to an explicit expression for the distribution function $f(R^{2},t)$ in the form
\begin{equation} \label{eq:18}
f(R^{2},t) = \frac{I_{0}}{2Da}\exp\left[- 2^{5/2}\frac{4\pi a^{1/2}}{15}\Gamma I_{0}D^{3/2} \left(t-\frac{R^{2}}{2Da}\right)^{5/2}\right] 
\end{equation}
It is recognized that the nucleation stage ends to the moment of time $t=t_{1}$ when the nucleation rate drops by $e$ times compared to its initial value.
Then, as follows from Eq.\eqref{eq:4},
\begin{equation} \label{eq:19}
\varphi(t_{1}) = \frac{1}{\Gamma} 
\end{equation}
With $\Gamma = 48.75$ we have from Eq.\eqref{eq:19} $\varphi(t_{1}) = 2.05\cdot10^{-2}$, i.e., the relative drop of the vapor supersaturation to the end of the
nucleation stage is about two percent, and Eq.\eqref{eq:3} certainly holds. Substituting Eq.\eqref{eq:17} into \eqref{eq:19} yields the expression time $t_{1}$
of duration of the nucleation stage,
\begin{equation} \label{eq:20}
t_{1} = \frac{1}{2}\left( \frac{15}{4\pi\Gamma a^{1/2}}\right)^{2/5}D^{-3/5}I_{0}^{-2/5}
\end{equation}
With the help of Eq.\eqref{eq:20}, the formula \eqref{eq:18} for distribution $f(R^{2},t)$ can be written in a more compact form as 
\begin{equation} \label{eq:21}
f(R^{2},t) = \frac{I_{0}}{2Da}\exp\left[-\left(\frac{R^{2}(t)-R^{2}}{R^{2}(t_{1})}\right)^{5/2}\right]
\end{equation}
The total number $N_{1} \equiv N(t_{1})$ of overcritical droplets formed during the nucleation stage is
\begin{equation} \label{eq:22}
N_{1} = \int_{0}^{R^{2}(t_{1})}f(R^{2},t_{1})dR^{2}
\end{equation}
Substitution of Eq.\eqref{eq:21} in Eq.\eqref{eq:22} leads to
\begin{equation} \label{eq:23}
N_{1} = \alpha_{d}I_{0}t_{1}
\end{equation}
where $\alpha_{d} \equiv \int_{0}^{1}\exp[-\xi^{5/2}] d\xi\approx 0.78$. With the help of Eq.\eqref{eq:20}, Eq.\eqref{eq:23} gives the relation
\begin{equation} \label{eq:24}
 N_{1} = \frac{\alpha_{d}}{2}\left(\frac{15}{4\pi\Gamma a^{1/2}}\right)^{2/5}\left(\frac{I_{0}}{D}\right)^{3/5}
\end{equation}
which is known as "law of 3/5".

As follows from Eqs.\eqref{eq:13} and \eqref{eq:20}, the maximal squared radius of the droplets to the end of the nucleation stage is 
\begin{equation} \label{eq:25}
R^{2}_{max} \equiv R^{2}(t_{1}) = 2Dat_{1} = \left(\frac{15a^{2}}{4\pi\Gamma} \frac{D}{I_{0}}\right)^{2/5}
\end{equation}
The mean squared radius of the droplets to the end of the nucleation stage is
\begin{equation} \label{eq:26}
 <R^{2}>_{1} \equiv \frac{\int_{0}^{R^{2}(t_{1})}R^{2}f(R^{2},t_{1})dR^{2}}{\int_{0}^{R^{2}(t_{1})}f(R^{2},t_{1})dR^{2}}
\end{equation}
Substituting the expression \eqref{eq:21} in Eq.\eqref{eq:26} yields 
\begin{equation} \label{eq:27}
<R^{2}>_{1} = \left(1 - \frac{\beta_{d}}{\alpha_{d}}\right)R^{2}(t_{1}) 
\end{equation}
where $\beta \equiv  \int_{0}^{1}\xi\exp[-\xi^{5/2}] d\xi \approx 0.34$. Thus, the relation between the mean squared radius and
the maximal squared radius of the droplets to the end of the nucleation stage can be written as
\begin{equation} \label{eq:28}
<R^{2}>_{1} = 0.56 \cdot R^{2}(t_{1})
\end{equation}
The last expression finalizes our program to link main characteristics of the final state of droplets to the end of the nucleation stage in the mean-field
approximation for vapor concentration. 

\section[The excluded volume approach.]{Description of the nucleation stage based on the excluded volume approach.}
The present approach is based on the fact that decrease of vapor supersaturation and respective decrease of the nucleation intensity are essentially
inhomogeneous over volume of gas-vapor mixture.  Inside the diffusion shells surrounding the previously nucleated droplets, the birth of new
droplets is practically suppressed, while it stays at the initial level in the remaining volume of mixture. As a consequence one may consider a certain volume
as excluded because the total volume of diffusion spheres with droplets at the center is eliminated from nucleation process. At the same time the influence on the
dynamics of growth of each droplet and on the profile of vapor concentration near droplet remains quite weak during the whole nucleation stage.

Excluded volume $V_{ex}(t)$ surrounding a single growing droplet of radius $R(t)$ can be defined from the condition that the total number of droplets nucleated per
unit of time around this particular droplet in sufficiently large volume $V$ of vapor-gas mixture at the current field of the nucleation rate is equal to the number
of droplets nucleated outside of excluded volume at the initial nucleation rate \cite{ref-6}. Suggesting that the nucleation rate $I_{0}$ remains unchanged (at the initial vapor
supersaturation) out of excluded volume, we have
\begin{equation} \label{eq:29}
\int_{V}d\overrightarrow{r}I(\zeta(r,t)) = I_{0}(V-V_{ex}(t))
\end{equation}
where the current local supersaturation $\zeta(r,t)$ of vapor \ignore{at the distance $r$ from the center of growing droplet}can be defined through the local concentration of
vapor $n(r,t)$ as 
\begin{equation} \label{eq:30}
 \zeta(r,t) = \frac{n(r,t) - n_{\infty}}{n_{\infty}}
\end{equation}
Taking into account a spherical symmetry of the problem, we arrive from \eqref{eq:29} to expression
\begin{equation} \label{eq:31}
V_{ex}(t) = 4\pi\int_{R(t)}^{\infty}\frac{I_{0}-I(\zeta(r,t))}{I_{0}}r^{2}dr
\end{equation}
Evidently the upper limit of the integral in Eq.\eqref{eq:31} could not exceed the mean distance between droplets and has been set as infinity formally, because the integrand
goes to zero quite fast.

To find value of excluded volume, one need to know an explicit profile of vapor concentration $n(r,t)$ in the space around the growing droplet. Further we apply
the results of Ref.\cite{ref-16} where the growth of single droplet was described on the base of nonstationary diffision equation with convection term arising from the motion
of vapor-gas mixture due to the movement of the surface of growing droplet. We neglect non-isothermal effects and the effect of the Stefan flow of vapor-gas medium
as well. Hence we get the following equation for the field of concentration around the droplet
\begin{equation} \label{eq:32}
\frac{\partial n(r,t)}{\partial t} = \frac{D}{r^{2}} \frac{\partial}{\partial r}\left[r^{2}\frac{\partial n(r,t)}{\partial r}\right] - \frac{R^{2}(t)}{r^{2}}\frac{dR(t)}{dt}\frac{\partial n(r,t)}{\partial r}
\end{equation}
In terms of self-similar variable $\rho$ defined as
\begin{equation} \label{eq:33}
 \rho = \frac{r}{R(t)}
\end{equation}
differential equation \eqref{eq:32} could be written as
\begin{equation} \label{eq:34}
\frac{d^{2}n(\rho)}{d\rho^{2}} + \left[\frac{2}{\rho}+\frac{R}{D}\frac{dR}{dt}\left(\rho - \frac{1}{\rho^{2}}\right)\right]\frac{dn(\rho)}{d\rho} = 0
\end{equation}
From the condition of  balance of number of vapor molecules on the surface of droplet
\begin{equation} \label{eq:35}
\frac{d}{dt}\left(\frac{4\pi}{3}n_{l}R^{3}\right) = \left(4\pi r^{2} D \frac{\partial n(r,t)}{\partial r}\right)\left.\vert\right._{r=R}
\end{equation}
we self-consistently obtain the expression
\begin{equation} \label{eq:36}
\frac{R}{D}\frac{dR}{dt} = \frac{1}{n_{l}}\frac{dn_{\rho}}{d\rho}|_{\rho=1} \equiv b
\end{equation}
or the rate of droplet’s radius growth in time. As a result of substitution of \eqref{eq:36} into \eqref{eq:32}, we come to an ordinary differential equation
\begin{equation} \label{eq:37}
\frac{d^{2}n(\rho)}{d\rho^{2}} + \left[\frac{2}{\rho}+b\left(\rho - \frac{1}{\rho^{2}}\right)\right]\frac{dn(\rho)}{d\rho} = 0
\end{equation}
for the profile $n(\rho)$ of vapor concentration with boundary conditions at the surface of droplet and infinitely far away from it in the form
\begin{equation} \label{eq:38}
n(\rho)|_{\rho=1} =  n_{\infty}
\end{equation}
\begin{equation} \label{eq:39}
n(\rho)|_{\rho=\infty} = n_{0}
\end{equation}
The self-similar solution of Eq.\eqref{eq:37} with boundary condition \eqref{eq:38} is
\begin{equation} \label{eq:40}
n(\rho) = n_{\infty} + bn_{l}\int_{1}^{\rho}\frac{dx}{x^{2}}\exp\left[-\frac{bx^{2}}{2}-\frac{b}{x}+\frac{3b}{2}\right]
\end{equation}
Substitution of solution \eqref{eq:40} into boundary condition \eqref{eq:39} results in the following transcendental equation on parameter $b$
\begin{equation} \label{eq:41}
b\int_{1}^{\infty}\frac{dx}{x^{2}}\exp\left[-\frac{bx^{2}}{2}-\frac{b}{x}+\frac{3b}{2}\right] = a
\end{equation}
where $a$ is the same dimensionless parameter as defined by Eq.\eqref{eq:12}. Under the condition
\begin{equation} \label{eq:42}
a^{1/2} \ll 1
\end{equation}
equation \eqref{eq:41} has a simple analytical solution
\begin{equation} \label{eq:43}
b = a(1 + \sqrt{\pi a /2})
\end{equation}
Choosing values $\zeta_{0} = 4$, $n_{l} = 3.3\cdot10^{22}$ cm$^{-3}$, $n_{\infty} = 1.3\cdot10^{17}$ cm$^{-3}$ we obtain $a = 1.58\cdot 10^{-5}$.
Thus the inequality \eqref{eq:42} is certainly valid, and self-similar solution \eqref{eq:40} can be presented with a sufficiently high accuracy as
\begin{equation} \label{eq:44}
 n(\rho) = n_{\infty} + (n_{0} - n_{\infty})\frac{\int_{1}^{\rho}\frac{dx}{x^{2}}\exp\left[-\frac{bx^{2}}{2}\right]}{\int_{1}^{\infty}\frac{dx}{x^{2}}\exp\left[-\frac{bx^{2}}{2}\right]}
\end{equation}
Using Eqs.\eqref{eq:44}, \eqref{eq:30} and \eqref{eq:2}, we obtain expressions for local supersaturation $\zeta(\rho)$ and relative decline $\varphi(\rho)$ of vapor 
supersaturation around the droplet in the form
\begin{equation} \label{eq:45}
\zeta(\rho) = \zeta_{0} \frac{\int_{1}^{\rho}\frac{dx}{x^{2}}\exp\left[-\frac{bx^{2}}{2}\right]}{\int_{1}^{\infty}\frac{dx}{x^{2}}\exp\left[-\frac{bx^{2}}{2}\right]}
\end{equation}
\begin{equation} \label{eq:46}
\varphi(\rho) = 1 - \frac{\int_{1}^{\rho}\frac{dx}{x^{2}}\exp\left[-\frac{bx^{2}}{2}\right]}{\int_{1}^{\infty}\frac{dx}{x^{2}}\exp\left[-\frac{bx^{2}}{2}\right]}
\end{equation}
As follows from Eqs.\eqref{eq:31} and \eqref{eq:33}, we get for self-similar regime of droplet growth
\begin{equation} \label{eq:47}
V_{ex} (t) = 4\pi R^{3}(t)\int_{1}^{\infty}\frac{I_{0} - I(\zeta(\rho))}{I_{0}}\rho^{2}d\rho
\end{equation}
The last expression could be rewritten as 
\begin{equation} \label{eq:48}
V_{ex}(t) = qV_{R}(t)
\end{equation}
where $V_{R} = \frac{4\pi}{3}R^{3}(t)$ is a droplet volume and $q$ is a ratio of excluded volume and droplet volume. As follows from Eqs.\eqref{eq:47} and 
\eqref{eq:48}, 
\begin{equation} \label{eq:49}
 q \equiv 3 \int_{1}^{\infty}d\rho\rho^{2}\left(1 - \frac{I(\zeta(\rho))}{I_{0}}\right)
\end{equation}
and $q$  does not depend on the size of droplet. This independence of size is provided by self-similarity of the vapor profile \eqref{eq:44}. The analogous parameter
introduced in Refs.\cite{ref-6,ref-7} depends on droplet size.

Let us notice that similar formalism was used in Ref.\cite{ref-17} for describing the kinetics of nucleation in binary glasses, where the concept of excluded zone with the size
defined by the expression similar to Eq. \eqref{eq:49} was also introduced. Expression \eqref{eq:48} for excluded volume $V_{ex}(t)$ was also used to describe the stage of nucleation
of bubbles in gas-supersaturated solution \cite{ref-12,ref-13}. However the expression for parameter $q$ was different from the present case.

To obtain $q$ with sufficiently high accuracy, we may use Eq.\eqref{eq:4} in the integrand of Eq.\eqref{eq:49}. Substituting Eq.\eqref{eq:4} into Eq.\eqref{eq:49}, 
integrating by parts and taking into account Eq.\eqref{eq:46}, we get
\begin{equation} \label{eq:50}
q = -1 + \frac{\Gamma}{\int_{1}^{\infty}he \frac{dx}{x^{2}}\exp\left[-\frac{bx^{2}}{2}\right]}\int_{1}^{\infty}dx \exp\left[-bx - \Gamma\varphi(\sqrt{2x})\right]
\end{equation}
While the strong inequality \eqref{eq:42} is fulfilled, the relation $\int_{1}^{\infty}\frac{dx}{x^{2}}\exp\left[-\frac{bx^{2}}{2}\right] \simeq 1$ is valid. According
to \eqref{eq:46}, we have $\varphi(\rho) \simeq  \frac{1}{\rho}$ at $\rho \ll \sqrt{2/b}$. Then parameter $q$ can be estimated with high accuracy at 
$\frac{1}{2}\Gamma^{2/3}b^{1/3} \ll 1$ as $q \approx \sqrt{2/b}$. Taking at $\zeta_{0} = 4$ the values $\Gamma = 48.75$ and $a = 1.58\cdot 10^{-5}$, we find 
$q \leq 3\cdot 10^{6}$. The exact result is $q = 2.89\cdot 10^{6}$.

Since ratio $q$ does not depend on the size of droplet, then the same ratio of the volumes is valid for the whole ensemble of overcritical droplets. In other words,
if the total volume of droplets at time $t$ equal to $V_{l}$ nucleation is suppressed in volume $V_{ex}^{tot} = qV_{l}$. Let $V$ be a total volume of system. Then
volume $V_{1}$, where the initial nucleation rate is kept, can be written as 
\begin{equation} \label{eq:51}
V_{1}(t) = V - V_{ex}^{tot}(t)
\end{equation}
The number $dN(\tau)$ of droplets nucleated \ignore{in the unit of volume}between moments of time $\tau$ and $\tau+d\tau$ is equal to
\begin{equation} \label{eq:52}
dN(\tau) = I_{0}\frac{V_{1}(\tau)}{V}d\tau
\end{equation}
Assuming radius of droplet at the time of its birth to be zero, using Eqs.\eqref{eq:36} and \eqref{eq:43} for the rate of droplet’s radius growth in time in the form
$dR^{2}/dt = 2Db$, one can arrive from \eqref{eq:51} to the following integral equation 
\begin{equation} \label{eq:53}
V_{1}(t) = V - qI_{0}\int_{0}^{t}dt_{1}V_{R}(t-\tau)V_{1}(\tau)
\end{equation}
where
\begin{equation} \label{eq:54}
V_{R}(t-\tau) = \frac{4\pi}{3}R^{3}(t-\tau) = \frac{4\pi}{3}(2Db(t-\tau))^{3/2}
\end{equation}
The integral equation \eqref{eq:53} is similar to equation for free volume obtained in Ref.\cite{ref-8}. However the analogue of parameter $q$ used in \cite{ref-8} has
been estimated with the help of solution of diffusion equation in the form of point sink with the sink intensity $dR^{2}/dt = 2Da$ instead of self-similar solution
\eqref{eq:44} with the rate of the droplet growth $dR^{2}/dt = 2Db$ as a consequence of droplet surface movement.

Introducing the relative part $z(t) \equiv \frac{V_{1}(t)}{V}$ of the volume where the initial nucleation rate is kept, we transform Eq.\eqref{eq:53} into integral
equation
\begin{equation} \label{eq:55}
z(t) = 1 - q\lambda t^{5/2}\int_{0}^{1}ds (1-s)^{3/2}z(ts)
\end{equation}
where
\begin{equation} \label{eq:56}
\lambda \equiv \frac{4\pi}{3}I_{0}(2Db)^{3/2}
\end{equation}
is a new parameter. The solution of Eq.\eqref{eq:55} can be obtained as a series \cite{ref-12,ref-13}
\begin{equation} \label{eq:57}
z(t) = \sum_{k=0}^{\infty}(-1)^{k}\frac{[q\lambda\cdot\Gamma(5/2)t^{5/2}]^{k}}{\Gamma\left(\frac{5k}{2}\right)}
\end{equation}
where $\Gamma(5/2)$ and $\Gamma(\frac{5k}{2}+1)$ are gamma functions. We are interested in time interval $0 \leq t \leq t_{1}$, where $t_{1}$ is the least positive
solution of equation $z(t) = 0$. At time $t_{1}$, the diffusion shells where nucleation of new droplets is suppressed fill the whole volume $V$ of a system. Thus 
$t_{1}$ corresponds to the end of nucleation stage. Since the series \eqref{eq:57} is converged quite fast, we may keep only first two terms of the series. As a result,
we obtain \cite{ref-12,ref-13}
\begin{equation} \label{eq:58}
 z(t) = 1 - \frac{2}{5}q\lambda t^{5/2}
\end{equation}
Thus duration of the nucleation stage equals to
\begin{equation} \label{eq:59}
t_{1} = \left(\frac{5}{2q\lambda}\right)^{2/5}
\end{equation}
and expression for $z(t)$ can be rewritten as
\begin{equation} \label{eq:60}
 z(t) = 1 - \left(\frac{t}{t_{1}}\right)^{5/2}
\end{equation}
Substituting expression \eqref{eq:56} for parameter $\lambda$ into Eq.\eqref{eq:59}, we find
\begin{equation} \label{eq:61}
t_{1} = \frac{1}{2}\left(\frac{15}{4\pi q}\right)^{2/5}\cdot\left(\frac{1}{b}\right)^{3/5}D^{-3/5}I_{0}^{-2/5}
\end{equation}
The number of nucleated droplets per unit of volume at time $t$ is determined as
\begin{equation} \label{eq:62}
N(t) = I_{0}\int_{0}^{t}z(t')dt' 
\end{equation}
Using expression \eqref{eq:58} for $z(t)$ yields
\begin{equation} \label{eq:63}
N(t) = I_{0}t\left[1-\frac{2}{7}\left(\frac{t}{t_{1}}\right)^{5/2}\right]
\end{equation}
Thus we obtain at the moment $t=t_{1}$ when the stage is over that
\begin{equation} \label{eq:64}
N_{1} \equiv N(t_{1}) = \frac{5}{7}I_{0}t_{1}
\end{equation}
Taking into account expression \eqref{eq:61} we get
\begin{equation} \label{eq:65}
N_{1} = \frac{5}{7}\left(\frac{15}{8\pi q}\right)^{2/5}\cdot\left(\frac{1}{b}\right)^{3/5}D^{-3/5}I_{0}^{3/5}
\end{equation}
For the maximal radius of droplet at the moment $t$ according to equation $dR^{2}/dt = 2Db$ we have
\begin{equation} \label{eq:66}
 R_{max}^{2} \equiv R^{2}(t_{1}) = 2Dbt_{1}
\end{equation}
and substituting \eqref{eq:61} gives
\begin{equation} \label{eq:67}
R_{max}^{2} = \left(\frac{15b D}{4\pi q I_{0}}\right)^{2/5}
\end{equation}
Now let us rewrite the integral relation \eqref{eq:62} by using Eqs.\eqref{eq:60} and \eqref{eq:54} and substituting the integration variable $\tau$ by $t-t'$ where $t'$ is a time in which dropet grow from zero size to 
$R'=R(t')$. Then again using equation $dR^{2}/dt = 2Db$, let us pass to a new integration variable $R'^{2}$ instead of $t'$. 
As a result we get
\begin{equation} \label{eq:68}
N(t) = \frac{I_{0}}{2Db}\int_{0}^{2Dbt}dR'^{2}\left[1-\left(\frac{t}{t_{1}}-\frac{R'^{2}}{2Dbt_{1}}\right)^{5/2}\right]
\end{equation}
This allows us to conclude that the droplet size-distribution function $f(R^{2},t)$ is
\begin{equation} \label{eq:69}
f(R^{2},t) = \frac{I_{0}}{2Db}\left[1-\left(\frac{t}{t_{1}}-\frac{R'^{2}}{2Dbt_{1}}\right)^{5/2}\right]\theta(t - R^{2}/(2Db))
\end{equation}
Substituting Eq.\eqref{eq:69} into Eq.\eqref{eq:26} and taking into account Eq.\eqref{eq:64}, one can derive for the average square radius of the droplet at the moment
of the end of the stage the following expression
\begin{equation} \label{eq:70}
<R^{2}>_{1} = \frac{7}{5}\int_{0}^{2Dbt_{1}}dR^{2}\frac{R^{2}}{2Dbt_{1}}\left[1-\left(1-\frac{R^{2}}{2Dbt_{1}}\right)^{5/2}\right]
\end{equation}
Passing to the new integration variable $z = \frac{R^{2}}{2Dbt_{1}}$ and taking into account Eq.\eqref{eq:66}, we obtain
\begin{equation} \label{eq:71}
<R^{2}>_{1} = \frac{7}{5}R^{2}(t_{1})\int_{0}^{1}dzz[1-(1-z)^{5/2}]
\end{equation}
Performing integration in Eq.\eqref{eq:71}, we get
\begin{equation} \label{eq:72}
<R^{2}>_{1} = \frac{11}{18}R^{2}(t_{1}) 
\end{equation}
As follows from the balance equation of the condensing matter 
\begin{equation} \label{eq:73}
 V(n_{0} - \widetilde{n}(t)) = n_{l}V_{l}(t)
\end{equation}
the averaged over volume of a system vapor concentration $ \widetilde{n}(t)$ can be determined as 
\begin{equation} \label{eq:74}
\widetilde{n}(t) = n_{0} -n_{l}\frac{V_{l}(t)}{V}
\end{equation}
Correspondingly, the average relative decline $\widetilde{\varphi}(t)$ of vapor supersaturation is
\begin{equation} \label{eq:75}
\widetilde{\varphi}(t) = \frac{1}{a}\frac{V_{l}(t)}{V} 
\end{equation}
At the moment $t_{1}$ of the end of the nucleation stage, the total excluded volume $V_{ex}^{tot}$ is equal to the whole volume of a system
\begin{equation} \label{eq:76}
V_{ex}^{tot} (t_{1}) = qV_{l}(t_{1}) = V
\end{equation}
and substituting Eq.\eqref{eq:76} into Eq.\eqref{eq:75} gives
\begin{equation} \label{eq:77}
\widetilde{\varphi}(t_{1}) = \frac{1}{aq} 
\end{equation}
For $a = 1.58\cdot10^{-5}$, $q = 2.89\cdot 10^{6} $ we get $\widetilde{\varphi}(t_{1}) = 2.2\cdot10^{-2}$. This estimate finalizes our program to link main characteristics of the final state of droplets to the end of the nucleation stage in the  excluded
volume approach.
\section{Discussion of the results.}
Let us compare the results of the excluded volume approach and the theory of mean-field vapor supersaturation for key characteristics at the end of the 
nucleation stage. We denote the quantities found in the mean field and exluded volume theories by upper indicies mf and ex, correspondigly. According to Eqs.\eqref{eq:20}
and \eqref{eq:61}, \eqref{eq:24} and  \eqref{eq:65}, \eqref{eq:28} and \eqref{eq:72}, we have the following ratios for the time $t_{1}$ of the duration of the nucleation stage, the total number $N_{1}$
of nucleated droplets, the maximal square radius $R_{max}^{2}$ and mean square radius $<R^{2}>_{1}$ of droplets 
\begin{equation} \label{eq:78}
\frac{t_{1}^{mf}}{t_{1}^{ex}} = \left(\frac{b}{a}\right)^{1/5}\left(\frac{qb}{\Gamma}\right)^{2/5}
\end{equation}
\begin{equation} \label{eq:79}
\frac{N_{1}^{mf}}{N_{1}^{ex}} = 1.09\cdot\left(\frac{b}{a}\right)^{1/5}\left(\frac{qb}{\Gamma}\right)^{2/5}
\end{equation}
\begin{equation} \label{eq:80}
\frac{(R_{max}^{2})^{mf}}{(R_{max}^{2})^{ex}} = \left(\frac{a}{b}\right)^{4/5}\left(\frac{qb}{\Gamma}\right)^{2/5}
\end{equation}
\begin{equation} \label{eq:81}
\frac{<R^{2}>_{1}^{mf}}{<R^{2}>_{1}^{ex}} = 0.92\cdot\left(\frac{a}{b}\right)^{4/5}\left(\frac{qb}{\Gamma}\right)^{2/5}
\end{equation}
In the case when the strong inequality $a^{1/2} \ll 1$ holds which provides the quasi-steady diffusion droplet growth, the parameters $b \approx a$, $q \approx \Gamma/a$,
and the right-hand sides of Eqs.\eqref{eq:78}-\eqref{eq:81} are close to unity. Thus both approaches give almost the same values of the quantities characterizing the
behavior of nucleating system on the nucleation stage. This proximity of the results is a consequence of smallness of the average value of the supersaturation decline
$\widetilde{\varphi}(t)$ at $a \ll 1$ in diffusion layers surrounding the droplets which was also noted and used in Ref.\cite{ref-8}. With increasing the temperature
of the system, the inequality $a^{1/2} \ll 1$ weakens, and the results of two approaches can differ more significantly.
\section{Conclusions.}
We applied the excluded-volume approach to the kinetics of nucleation in supersaturated vapor-gas mixture. It allows us to properly describe the physics of the process by 
taking into account the vapor concentration inhomogeneities arising from the depletion of vapor due to non-stationary diffusion onto growing droplets. In contrast to 
other works we used an exact unsteady self-similar solution of the non-stationary diffusion equation with convection term describing the flow of the gas-vapor mixture
caused by moving surface of single growing droplet. We have shown that for this system the strong inequality $a^{1/2} \ll 1$ for the non-stationary parameter $a$ holds.
Obtained results were compared with the traditional mean-field approach when vapor supersaturation is decreasing uniformly over volume of the system. It was shown that
both approaches give almost the same values of the quantities characterizing the behavior of nucleating system on the nucleation stage while the strong inequality 
$a^{1/2} \ll 1$ holds but may differ with increase of temperature.
\section*{Acknowledgements.}
This work was supported by the Russian Foundation for Basic Research (grant 13-03-01049-a) and the Program of Development of St.Petersburg State University (grant 0.37.138.2011). 
Maxim Markov is thankful for the DRE grant of Ecole Polytechnique de Paris, France. 


\begin{thebibliography}{1}
\bibitem{ref-1}
N.N. Tunitskij, Zh. Fiz. Khimii 15 (1941) 10.
\bibitem{ref-2}
H. Wakeshima, Phys.Soc.Japan 9 (1954) 400.
\bibitem{ref-3}
F.M. Kuni, Problemy kinetiki kondensacii,Preprint ITF-83-79R, Kiev, 1983.
\bibitem{ref-4}
F.M. Kuni, A.P. Grinin, A.S. Kabanov, Colloid J. of the USSR 45 (1983) 385. 
\bibitem{ref-5}
V.V. Slezov, Kinetics of first-order phase transitions, Wiley-VCH, Berlin, 2009.
\bibitem{ref-6}
A.J. Pesthy, R.C. Flagan, J.H. Seinfeld, J. of Colloid and Interface Science, 82 (1981) 465. 
\bibitem{ref-7}
A.J. Pesthy, R.C. Flagan, J.H. Seinfeld, J. of Colloid and Interface Science, 91 (1983) 525.
\bibitem{ref-8}
V.  Kurasov, Physica A 226 (1996) 117. 
\bibitem{ref-9}
V. Kurasov, Phys.Rev.E 63 (2001) 056123.
\bibitem{ref-10}
A. P. Grinin, I. A. Zhuvikina, F. M. Kuni, Colloid Journal 66 (2004) 277.
\bibitem{ref-11}
A. P. Grinin, I. A. Zhuvikina, F. M. Kuni, H. Reiss, J.Chem.Phys. 121 (2004) 12490
\bibitem{ref-12}
A.E. Kuchma, F.M. Kuni, A.K. Shchekin, Phys.Rev.E. 80 (2009) 061125. 
\bibitem{ref-13}
A.E. Kuchma, F.M. Kuni, A.K. Shchekin, Vestnik St.Petersburg Univ. 4(4) (2009) 321. 
\bibitem{ref-14}
Ya.B. Zeldovich, Sov. Phys. JETP, 12 (1942) 325.
\bibitem{ref-15}
N.A. Fuchs, Evaporation and Droplet Growth in Gaseous Media, Pergamon, London, 1959.
\bibitem{ref-16}
A.E. Kuchma, A.K. Shchekin, Colloid Journal 74 (2012) 215. 
\bibitem{ref-17}
M.P. Shepilov, Journal of Non-Crystalline Solids 208 (1996) 64.
\end{thebibliography}
\end{document}